\documentclass{PoS}
\usepackage{amsmath,amssymb, bm}

\newcommand{\rmd}{{\rm d}}

\title{The perturbative Pomeron with NLO accuracy: Jet-Gap-Jet Observables\thanks{In collaboration with B.~Murdaca, J.~D.~Madrigal, A.~Sabio Vera}}

\ShortTitle{The perturbative Pomeron with NLO accuracy: Jet-Gap-Jet Observables}

\author{\speaker{Martin Hentschinski}\\
Instituto de Ciencias Nucleares \\
Universidad Nacional Aut\'onoma de M\'exico\\
Apartado Postal 70-543 \\
M\'exico D.F. 04510 MX      
 \\
        E-mail: \email{hentschinski@correo.nucleares.unam.mx}}

      \abstract{We give an overview of  the calculation of the forward jet vertex associated to a rapidity gap (coupling of a hard pomeron to a jet) in the Balitsky-Fadin-Kuraev-Lipatov (BFKL) formalism at next-to-leading order (NLO). This result  allows, together with the NLO non-forward gluon Green function, to perform NLO studies of jet production in diffractive events (Mueller-Tang dijets, as a well-known example).}

\FullConference{XXIII International Workshop on Deep-Inelastic Scattering,\\
		27 April - May 1 2015\\
		Dallas, Texas}

\begin{document}

\section{Introduction}
\label{sec:intro}

Understanding of the phenomenological very successful results of
Regge-theory from first principles remains till today one of the big
open questions of nuclear and particle physics. Within this framework
$t$-channel exchanges of vacuum quantum number are described by a
special kind of Regge trajectory, the Pomeron. It is not only
responsible for the asymptotic growth of inclusive cross-sections
(associated with a high multiplicity final state), but also governs
the high energy behavior of diffractive cross-sections (associated
with a low multiplicity final state).  Within QCD perturbation theory,
the Pomeron is obtained from high energy factorization and
corresponding resummation of large logarithms in the center of mass
energy, as provided by the famous BFKL result \cite{BFKL}.  For
diffractive events, which require the non-forward BFKL Pomeron at
finite momentum transfer $t$, the BFKL Green's function, which
achieves the high energy resummation, has been calculated up to
next-to-leading logarithmic (NLL) accuracy in \cite{nonforward}.
While the bulk of diffractive cross-sections is of non-perturbative
nature, it is possible to identify kinematic configurations with a
hard scale which are in principle accessible to a perturbative
treatment.  A possible observable, original proposed in \cite{tang},
to test the BFKL Green's function for finite $t$ is given by
forward-backward jets separated by a large empty region in the
detector, a so-called rapidity gap. A complete description of such
events at NLL requires apart from the Green's function also the
couplings of the Pomeron to the jet at next-to-leading order
(NLO). While virtual corrections can be extracted from \cite{fadin},
real corrections have been calculated recently in \cite{mtvnlo}.  This
calculation employs Lipatov's high energy effective action
\cite{effective}, making use of a framework for higher order
corrections within this action which has been developed and tested in
\cite{ourEff}. In the following we present our final result for the
jet vertex for jets with rapidity gap within collinear
factorization. For details we refer the interested reader to
\cite{mtvnlo}.

\section{The NLO Mueller-Tang Jet Vertex}
To define an infrared and collinear safe jet cross sections at NLO, it is necessary to convolute the partonic cross 
section with a jet function $S_J$:
\begin{equation}
\frac{\rmd\hat{\sigma}_J}{\rmd J_1\rmd J_2\rmd^2\bm{k}}=\rmd\hat{\sigma}\otimes S_{J_1}S_{J_2},\quad \rmd J_i=\rmd^{2+2\epsilon}\bm{k}_{J_i}\rmd y_{J_i},\,i=1,2.
\end{equation}
Infrared finiteness imposes general constraints on the jet function  \cite{seymour}. For two final state partons, the  jet function $S_J^{(3)}(\bm{p},\bm{q},zx,x)$ must be $ \{ \bm{q}, z \} \leftrightarrow \{\bm{p}, 1-z\}$ symmetric, and must reduce to the one final state parton distribution  $S_{J}^{(2)}(\bm{p},x)=x\,\delta\left(x-\frac{|\bm{k}_J|e^{y_J}}{\sqrt{s}}\right)\delta^{2+2\epsilon}(\bm{p}-\bm{k}_J)$ in the soft and collinear limits. In particular
\begin{equation}\label{propjet}
S^{(3)}_J(\bm{p},\bm{q},zx,x)\stackrel{\bm{p}\to 0}{\longrightarrow}S^{(2)}_J(\bm{k},zx);\quad S^{(3)}_J(\bm{p},\bm{q},zx,x)\stackrel{\tfrac{\bm{q}}{z}\to\tfrac{\bm{p}}{1-z}}{\longrightarrow}S_J^{(2)}(\bm{k},x).
\end{equation}
Adding to our result  the virtual  corrections calculated in \cite{fadin}, as well as corresponding   UV renormalization of the QCD Lagrangian, and absorbing initial state collinear emissions into a redefinition of parton distribution functions, we obtain,  
\begin{equation}
\begin{aligned}
\hspace{-0.25cm}\frac{\rmd\sigma_{J,H_1H_2}}{\rmd J_1\rmd J_2\rmd^2\bm{k}}&=\frac{1}{\pi^2}\int\rmd\bm{l}_1\rmd\bm{l}_1'\rmd\bm{l}_2\rmd\bm{l}_2'\frac{\rmd V(\bm{l}_1,\bm{l}_2,\bm{k},\bm{p}_{J,1},y_1,s_0)}{\rmd J_1}
\\\times
&
 G\left(\bm{l}_1,\bm{l}_1',\bm{k}, \frac{\hat{s}}{s_0}\right)
G\left(\bm{l}_2,\bm{l}_2',\bm{k},\frac{\hat{s}}{s_0} \right)\frac{\rmd V(\bm{l}_1',\bm{l}_2',\bm{k},\bm{p}_{J,2},y_2,s_0)}{\rmd J_2},
\end{aligned}
\end{equation}
where $\hat{s}=x_1x_2s$, $x_0= -t / (M_{x, \rm max}^2 - t)$ and
\begin{align}\label{jetv}
&\frac{\rmd V}{\rmd J}=\sum_{j=\{q_k,\bar{q}_k,g\}}^{k=1,\cdots,n_f}\int_{x_0}^1\rmd x\,f_{j/H}(x,\mu_F^2)
\left(\tfrac{\rmd \hat{V}^{(0)}_j}{\rmd J}+\tfrac{\rmd \hat{V}^{(1)}_j}{\rmd J}\right),
\, \, 
\frac{\rmd \hat{V}_j^{(0)}}{\rmd J} = \frac{\alpha_s^2 C_j^2 }{N_c^2 -1}S_J^{(2)}({\bm k}, x),
\notag\\
&
\frac{\rmd \hat{V}_j^{(1)}}{\rmd J}=
\int d\Gamma^{(2)} \left(\frac{\rmd \hat{V}_{j,\,v}^{(1)}}{\rmd J}+\frac{\rmd \hat{V}_{j,\,r}^{(1)}}{\rmd J}+\frac{\rmd \hat{V}_{j,\,{\rm UV\,ct.}}^{(1)}}{\rmd J}
+\frac{\rmd \hat{V}_{j,\,{\rm col.\,ct.}}^{(1)}}{\rmd J} \right),\notag\\
&\frac{\rmd\hat{V}^{(1)}_{r,\,\{q_k/\bar{q}_k,g\}}}{\rmd J}=\left\{h^{(1)}_{r,\,qqg}, h^{(1)}_{r,\,q\bar{q}g}+h^{(1)}_{r,\,ggg}\right\}S^{(3)}_J(\bm{p},\bm{q},zx,x),\\&\tfrac{\rmd \hat{V}_{\{q_k/\bar{q}_k,g\},\,{\rm UV\,ct.}}^{(1)}}{\rmd J}=\{h_q^{(0)},h_g^{(0)}\}\frac{\alpha_{s,\epsilon}}{2\pi}\frac{\beta_0}{\epsilon}S^{(2)}_J(\bm{k},x),~~\tfrac{\rmd\hat{V}_{\{g,q/\bar{q}\}}^{(0)}}{\rmd J}=h^{(0)}_{\{g,q\}}S_J^{(2)}(\bm{k},x),\notag\\&\tfrac{\rmd \hat{V}_{j,\,{\rm col.\,ct.}}^{(1)}}{\rmd J}=-\frac{\alpha_{s,\epsilon}}{2\pi}
\left(\tfrac{1}{\epsilon}+\ln\tfrac{\mu_F^2}{\mu^2}\right)
\int_0^1\rmd z\,S_J^{(2)}(\bm{k},zx)\sum_{i=\{q_\ell,\bar{q}_\ell,g\}}^{\ell=1,\cdots,n_f}h_i^{(0)}P_{ij}^{(0)}(z),\notag
\end{align}
with $\beta_0=\frac{11}{3}C_a-\frac{2}{3}n_f$, $P_{ij}^{(0)}(z)$ the
leading order DGLAP splitting functions and $C_{q,\bar{q}} = C_f, C_g = C_a$. The result for $\frac{\rmd
  \hat{V}^{(1)}_{j, v}}{\rmd J}$ can be extracted from \cite{fadin}.
Note that, if the result for the resummed jet cross-section is truncated at next-to-leading order in $\alpha_s$, our result is independent of the scale $s_0$ as required. To arrive at a physical representation of this vertex in dimension four we introduce a phase space slicing parameter, 
$\lambda^2\ll\bm{k}^2$, to regularize the singular regions in phase space. Using  the limits in  
Eq.~\eqref{propjet} we can rewrite $\rmd V_{q,g} / \rmd J$ in terms of $\lambda$~ and, 
introducing the notations ($i=1,2$)
\begin{align}
\label{pepe}
     P_{0}(z) =C_a&\big[\tfrac{2(1-z)}{z}+z(1-z)\big],\quad P_{1}(z)=C_a\big[\tfrac{2z}{[1-z]_+}+z(1-z)\big]  ,\notag\\
P_{qq}^{(0)}(z)  = C_f& \left(\frac{1+z^2}{1-z} \right)_+, \quad P_{qg}^{(0)(z)}= \frac{z^2 + (1-z)^2}{2} \; , \notag \\
 P^{(0)}_{gq}(z)= C_f& \frac{1 + (1-z)^2}{z}, ~~ P_{gg}^{(0)}(z) = P_0(z) + P_1(z) + \frac{\beta_0}{2}\delta(1-z) \; ,
\notag \\
    \alpha_s= \alpha_s (\mu^2) & , \qquad \phi_i=\arccos\tfrac{ \bm{l}_i \cdot ( \bm{k} - {\bm l}_i)}{|\bm{l}_i||\bm{k}-\bm{l}_i|},
\notag \\
    J_{1} ({\bm q}, {\bm k}, {\bm l}_i, z) & = \frac{1}{4} \bigg[ 2
    \frac{{\bm k}^2}{{\bm p}^2} \bigg(\frac{(1-z)^2}{{\bm \Delta}^2} -
    \frac{1}{{\bm q}^2} \bigg) - \frac{1}{{\bm \Sigma}_i^2} \bigg(
    \frac{({\bm l}_i - z {\bm k})^2}{{\bm \Delta}^2} - \frac{{\bm
        l}_i^2}{{\bm q}^2}
    \bigg)\notag \\
    & \qquad - \frac{1}{{\bm \Upsilon}_i^2} \bigg( \frac{({\bm l}_i -
      (1-z) {\bm k})^2}{{\bm \Delta}^2} - \frac{({\bm l}_i - {\bm
        k})^2}{{\bm q}^2} \bigg)
    \bigg],~i=1,2;  \notag\\
    J_{2} ({\bm q}, {\bm k}, {\bm l}_1, {\bm l}_2) &= \frac{1}{4}
    \bigg[ \frac{{\bm l}_1^2}{ {\bm p}^2 {\bm \Upsilon}^2_1} + \frac{(
      {\bm k} - {\bm l}_1)^2}{ {\bm p}^2 {\bm \Sigma}^2_1} +
    \frac{{\bm l}_2^2}{ {\bm p}^2 {\bm \Upsilon}^2_2} + \frac{( {\bm
        k} - {\bm l}_2)^2}{ {\bm p}^2 {\bm \Sigma}^2_2}
    \notag\\
    & \hspace{-1cm}- \frac{1}{2} \bigg( \frac{({\bm l}_1 - {\bm
        l}_2)^2}{{\bm \Sigma}_1^2 {\bm \Sigma}_2^2} + \frac{({\bm k} -
      {\bm l}_1 - {\bm l}_2)^2}{ {\bm \Upsilon}_1^2 {\bm \Sigma}_2^2 }
    + \frac{({\bm k} - {\bm l}_1 - {\bm l}_2)^2}{ {\bm \Sigma}_1^2
      {\bm \Upsilon}_2^2 } + \frac{({\bm l}_1 - {\bm l}_2)^2}{{\bm
        \Upsilon}_1^2 {\bm \Upsilon}_2^2} \bigg)
    \bigg],
\end{align}
we present our expression for those jets with a quark as the initial state, {\it i.e.}
\begin{align}
 &   \frac{\rmd\hat{V}^{(1)}_q(x, {\bm k}, {\bm l}_1, {\bm l}_2; x_J, {\bm k}_J; M_{X,\text{max}}, s_0)}{\rmd J}  =   v^{(0)}\frac{\alpha_s}{2 \pi}  \big(Q_1 + Q_2 + Q_3 \big)
\end{align}
\begin{align}
& Q_1 =   S_J^{(2)}({\bm k}, x) C_f^2 { \Bigg[-\frac{\beta_0}{4}
\bigg\{\left[\ln\left(\frac{\bm{l}_1^2}{\mu^2}\right)+\ln\left(\frac{(\bm{l}_1-\bm{k})^2}{\mu^2}\right)+\{1\leftrightarrow 2\}\right]} 
\notag \\ &  -\frac{20}{3}\bigg\} -4C_f + \frac{C_a}{2}\bigg(\bigg\{
\frac{3}{2\bm{k}^2}
\bigg[
\bm{l}_1^2\ln\left(\frac{(\bm{l}_1-\bm{k})^2}{\bm{l}_1^2}\right)+(\bm{l}_1-\bm{k})^2 \; \cdot
\notag \\
& \ln\left(\frac{\bm{l}_1^2}{(\bm{l}_1-\bm{k})^2}\right)  -4|\bm{l}_1||\bm{l}_1-\bm{k}|\phi_1\sin\phi_1\bigg]
  -\frac{3}{2}\bigg[\ln\left(\frac{\bm{l}_1^2}{\bm{k}^2}\right)
 \notag \\
&  +\ln\left(\frac{(\bm{l}_1-\bm{k})^2}{\bm{k}^2}\right)\bigg]
 -\ln\left(\frac{\bm{l}_1^2}{\bm{k}^2}\right)\ln\left(\frac{(\bm{l}_1-\bm{k})^2}{s_0}\right)
-\ln\left(\frac{(\bm{l}_1-\bm{k})^2}{\bm{k}^2}\right) \; \cdot
\notag \\
&
\ln\left(\frac{\bm{l}_1^2}{s_0}\right) -2\phi_1^2+\{1\leftrightarrow 2\}\bigg\} +2\pi^2+\frac{14}{3}\bigg)\Bigg]\;,
\end{align}
\begin{align}
& Q_2 =
\int_{z_0}^1 \rmd z  \,\, 
 S_J^{(2)}({\bm k}, z x)%
\bigg[ \ln \frac{\lambda^2}{\mu^2_F}   \left(  C_f^2
P^{(0)}_{qq}(z)+ {C_a^2}P^{(0)}_{gq}(z) \right) 
 \notag \\
&
+ C_f (1-z)\left(C_f^2 -\frac{2}{z} {C_a^2}\right)
+2 C_f (1+z^2)\left(\frac{\ln(1-z)}{1-z}\right)_+\bigg] \; ,
\end{align}
\begin{align}
& Q_3 = \int_0^1 \rmd z\int\frac{\rmd^2\bm{q}}{\pi} \bigg[
 \Theta\left(\hat{ M}^2_{X,{\rm max}}-\frac{({ \bm p } - z {\bm k})^2}{z(1-z)}\right) S_J^{(3)}(\bm{p},\bm{q},(1-z)x,x)  C_f^2  
\notag \\
&
  P^{(0)}_{qq}(z)\Theta\left(\frac{|\bm{q}|}{1-z}-\lambda^2\right)  
{ \frac{  \bm{k}^2}{\bm{q}^2(\bm{p}-z\bm{k})^2}}
 + \Theta\left(\hat{M}_{X, \rm max}^2 - \frac{{\bm \Delta}^2}{z(1-z)} \right)
 \notag \\
&  S_J^{(3)}(\bm{p},\bm{q},zx,x)  
   P_{gq}^{(0)}(z) 
 \big\{C_f C_a [J_1(\bm{q},\bm{k},\bm{l}_1)+J_1(\bm{q},\bm{k},\bm{l}_2)]
\notag \\ &
\hspace{6cm}
+ {C_a^2} J_2(\bm{q},\bm{k},\bm{l}_1,\bm{l}_2) \Theta(\bm{p}^2-\lambda^2)
    \big\}\bigg] \; .
\end{align}
In a similar way, the equivalent gluon-generated forward jet vertex reads
\begin{align}
\label{apocalipsis}
&\frac{\rmd\hat{V}^{(1)}(x,\bm{k},\bm{l}_1,\bm{l}_2;x_J,\bm{k}_J;M_{X,\,{\rm max}},s_0)}{\rmd J}  =v^{(0)}\frac{\alpha_s}{2\pi}\big( G_1 + G_2 + G_3 \Big)
\end{align}
\begin{align}
& G_1  =
C_a^2\,S_J^{(2)}(\bm{k},x)
\Bigg[C_a \left( \pi^2- \frac{5}{6} \right)- \beta_0 \left( \ln\frac{\lambda^2}{\mu^2} -  \frac{4}{3} \right) 
\notag \\
&
+\left(\frac{\beta_0}{4} + \frac{11 C_a}{12}+\frac{n_f}{6 C_a^2}\right)
\left( \ln\frac{\bm{k}^4}{ {\bm l}_1^2  ({\bm k} - {\bm l}_1^2)}
+
 \ln\frac{\bm{k}^4}{ {\bm l}_2^2  ({\bm k} - {\bm l}_2)^2 } 
\right)
\notag \\
&
+\frac{1}{2}\bigg\{C_a\bigg(
\ln^2\frac{\bm{l}_1^2}{(\bm{k}-\bm{l}_1)^2}
+
\ln\frac{\bm{k}^2}{\bm{l}_1^2}\ln\frac{\bm{l}_1^2}{s_0}
+
\ln\frac{\bm{k}^2}{(\bm{k}-\bm{l}_1)^2}\ln\frac{(\bm{k}-\bm{l}_1)^2}{s_0}
\bigg)
\notag\\
&-\bigg(\frac{n_f}{3 C_a^2} + \frac{11 C_a}{6}\bigg)
\frac{\bm{l}_1^2-(\bm{k}-\bm{l}_1)^2}{\bm{k}^2}
\ln\frac{\bm{l}_1^2}{(\bm{k}-\bm{l}_1)^2}
-2\bigg(\frac{n_f}{C_a^2}+4C_a\bigg)
\notag\\
&  \frac{(\bm{l}_1^2(\bm{k}-\bm{l}_1)^2)^{\frac{1}{2}}}{\bm{k}^2} \phi_1\sin\phi_1 
+\frac{1}{3}\bigg(
C_a+\frac{n_f}{C_a^2}
\bigg)\bigg[
16\frac{(\bm{l}_1^2(\bm{k}-\bm{l}_1)^2)^{\frac{3}{2}}}{(\bm{k}^2)^3}\phi_1\sin^3\phi_1
\notag\\
& -4\frac{\bm{l}_1^2(\bm{k}-\bm{l}_1)^2}{(\bm{k}^2)^2}  \bigg(
2-\frac{\bm{l}_1^2-(\bm{k}-\bm{l}_1)^2}{\bm{k}^2}\ln\frac{\bm{l}_1^2}{(\bm{k}-\bm{l}_1)^2}
\bigg)
\sin^2\phi_1+\frac{(\bm{l}_1^2(\bm{k}-\bm{l}_1)^2)^{\frac{1}{2}}}{(\bm{k}^2)^2}
\notag\\ & \cos\phi_1
\bigg(4\bm{k}^2 -12(\bm{l}_1^2(\bm{k}-\bm{l}_1)^2)^{\frac{1}{2}}\phi_1\sin\phi_1-(\bm{l}_1^2-(\bm{k}-\bm{l}_1)^2)\ln\frac{\bm{l}_1^2}{(\bm{k}-\bm{l}_1)^2}\bigg)
\bigg]
\notag\\&-2C_a\phi_1^2+\{\bm{l}_1\leftrightarrow\bm{l}_2,\phi_1\leftrightarrow\phi_2\}\bigg\}\Bigg]
\end{align}
\begin{align}
 & G_2 =   \int_{z_0}^1\rmd z\,S_J^{(2)}(\bm{k},zx) \bigg\{  2n_f P_{qg}^{(0)}(z)
\left(C_f^2 \ln \frac{\lambda^2}{\mu_F^2}  +  C_a^2 \ln(1-z) \right) 
\notag  
 \\ &
 + C_a^2 P_{gg}^{(0)}(z) \ln\frac{\lambda^2}{\mu_F^2}  + C_f^2 n_f + 2 C_a^3 z \bigg(  (1-z)\ln(1-z)   + 2 \left[\frac{\ln (1-z)}{1-z} \right]_+ \bigg)\bigg\}
\end{align}
\begin{align}
&G_3 = 
\int_0^1 \rmd z\int\frac{\rmd^2\bm{q}}{\pi} \bigg\{
n_f P^{(0)}_{qg}(z) 
\bigg[
C_a^2 \Theta\left(\hat{M}_{X,{\rm max}}^2-\frac{z {\bm p}^2}{(1-z)}\right)
    S^{(3)}_J(\bm{k}-z\bm{q},z\bm{q},zx,x)
\notag \\
&
 \bigg[ \frac{\Theta({\bm p^2 - \lambda^2}) {\bm k}^2}{({\bm p}^2 + {\bm q}^2) {\bm p}^2} +
\frac{ {\bm k}^2}{({\bm p}^2 + {\bm q}^2) {\bm q}^2} \bigg]
- \Theta\left(\hat{M}_{X,{\rm max}}^2-\frac{{\bm \Delta}^2}{z(1-z)}\right) 
  S^{(3)}_J(\bm{p},\bm{q}, zx, x)
\notag \\
&
\bigg(C_a^2  \frac{ {\bm k}^2}{({\bm p}^2 + {\bm q}^2) {\bm q}^2} 
-   
 2 C_f^2 
 \frac{{\bm k}^2 \Theta({\bm q}^2 - \lambda^2)}{({\bm p}^2 + {\bm q}^2) {\bm q}^2  } \bigg)\bigg]
+ P_1(z)  \Theta\left(\hat{ M}^2_{X,{\rm max}}-\frac{({ \bm p } - z {\bm k})^2}{z(1-z)}\right) 
\notag \\ & 
 S_J^{(3)}({\bm p}, {\bm q}, (1-z)x, x) \frac{(1-z)^2 {\bm k}^2}{(1-z)^2 ({\bm p} - z {\bm k})^2 + {\bm q}^2}
\bigg[  \Theta\left(\frac{|\bm{q}|}{1-z}-\lambda\right)  \frac{1}{{\bm q}^2} 
\notag \\
 &+ 
\Theta\left(\frac{|\bm{p} - z {\bm k}|}{1-z}-\lambda\right)  \frac{1}{({\bm p} - z {\bm k})^2}\bigg]
 + 
\Theta\left(\hat{M}_{X, \rm max}^2 - \frac{{\bm \Delta}^2}{z(1-z)} \right)
 S_J^{(3)}(\bm{p},\bm{q},zx,x) 
& \notag \\
 &
\bigg[
\frac{n_f }{ C_a^2}  P_{qg}^{(0)} \bigg(J_2(\bm{q},\bm{k},\bm{l}_1,\bm{l}_2)   - \frac{{\bm k}^2}{{\bm p}^2( {\bm q}^2 + {\bm p}^2)}\bigg)
- n_f  P_{qg}^{(0)} \bigg(
J_{1} ({\bm q}, {\bm k}, {\bm l}_1, z)  \notag \\
&  + J_{1} ({\bm q}, {\bm k}, {\bm l}_2, z)  
 \bigg) 
+   P_0(z) 
 \bigg( J_1(\bm{q},\bm{k},\bm{l}_1)+J_1(\bm{q},\bm{k},\bm{l}_2)  + J_2(\bm{q},\bm{k},\bm{l}_1,\bm{l}_2) \Theta(\bm{p}^2-\lambda^2)
    \bigg)\bigg] 
\bigg\}.
\end{align}
These expressions are in a form suitable for  phenomenological studies. It is important to note that its 
convolution with the nonforward BFKL Green function with exact treatment of the running of the QCD coupling 
is complicated. The use of Monte Carlo integration techniques~\cite{Chachamis:2011nz} appears therefore to be preferable  since they allow to generate exclusive distributions needed to 
describe different diffractive data in hadronic collisions. For more inclusive observables, analytic methods might be a  valuable alternative   where one might for a complete NLL treatment  follow the treatment proposed in \cite{Chirilli:2013kca,Ross:2008ur}.
\section*{Acknowledgments}
%
This work was supported in part by UNAM-DGAPA-PAPIIT grant number 101515
and CONACyT-Mexico grant number 128534.


\end{document}